\documentclass[apl,
a4paper,
showpacs,
twocolumn,
superscriptaddress]{revtex4-1}
\usepackage{amssymb}
\usepackage{amsmath}
\usepackage{amsfonts}
\usepackage{graphicx}
\usepackage{bm}
\usepackage{color}
\usepackage{multirow}
\usepackage{natbib}
\usepackage{hyperref}
\usepackage[normalem]{ulem}
\usepackage{verbatim}
\def \be {\begin{equation}}
\def \ee {\end{equation}}
\def \bea {\begin{align}}
\def \eea {\end{align}}
\def \p {\partial}
\def \BEA {\begin{eqnarray}}
\def \EEA {\end{eqnarray}}

\def \BC {\begin{cases}}
\def \EC {\end{cases}}

\def \be {\begin{equation}}
\def \ee {\end{equation}}
\def \bea {\begin{align}}
\def \eea {\end{align}}
\def \p {\partial}
\def \BEA {\begin{eqnarray}}
\def \EEA {\end{eqnarray}}

\def \BC {\begin{cases}}
\def \EC {\end{cases}}

\usepackage{graphicx}
\usepackage{dcolumn}
\usepackage{bm}
\usepackage{amssymb}
\usepackage{amsmath}
\usepackage{verbatim}
\usepackage{color}
\usepackage{ulem}
\def \be {\begin{equation}}
\def \ee {\end{equation}}
\def \bea {\begin{align}}
\def \eea {\end{align}}
\def \p {\partial}
\def \BEA {\begin{eqnarray}}
\def \EEA {\end{eqnarray}}

\def \BC {\begin{cases}}
\def \EC {\end{cases}}

\usepackage{epsf}
\usepackage{graphicx}
\def \e {\begin{equation}}
\def \en {\end{equation}}
\def \ea {\begin{align}}
\def \eea {\end{align}}
\def \p {\partial}
\def \BEA {\begin{eqnarray}}
\def \EEA {\end{eqnarray}}
\def \BC {\begin{cases}}
\def \EC {\end{cases}}

\begin{document}

\title{Plasmonic  Helicity-Driven  Detector  of Terahertz Radiation }
\author{ I.V.~Gorbenko}
\affiliation{A. F.~Ioffe Physico-Technical Institute,
194021 St.~Petersburg, Russia}

\author{ V. Yu.~Kachorovskii}
\affiliation{A. F.~Ioffe Physico-Technical Institute,
194021 St.~Petersburg, Russia}
\affiliation{L. D.~Landau Institute for Theoretical Physics, Kosygina
  street 2, 119334 Moscow, Russia}

\author{M. S.~Shur}
\affiliation{Rensselaer Polytechnic Institute, 110, 8$^{th}$
Street, Troy, NY, 12180, USA}

\date{\today}

\begin{abstract}
   We develop a theory of
    the helicity driven
        nolinear  dc response of
        gated two-dimensional  electron gas
     to   the terahertz radiation.
            We
            demonstrate that  the helicity-sensitive part of the response
                        dramatically increases in the vicinity of the plasmonic resonances and
                        oscillates with the phase shift between excitation signals on the source and drain.  The resonance line shape is
                        an
                        asymmetric function of the frequency deviation from the resonance.
      In contrast, the helicity-insensitive part of the  response is symmetrical.
                     These properties  yield  significant advantage for  using plasmonic detectors as  terahertz  and far infrared spectrometers and interferometers.
\end{abstract}
\maketitle

Plasma wave terahertz (THz) emitters \cite{02,03,04,05,11,14,ad3} and detectors
\cite{08,09,010,0015,3,2,6,12,013,014,0101,99,ad1,ad2,7,nep1,nep2} based on field effect transistors (FETs)  are promising candidates for filling the
famous THz gap. Although the emission of radiation requires some special  conditions, particularly, specific
boundary  conditions (BC) \cite{02}
 or the electron velocity exceeding the plasma velocity \cite{plasmonic}
 the detection only relies on the device nonlinearity \cite{08}.
 Impinging THz or sub-THz radiation excites plasma waves in the FET channel. Rectification of these waves
 leads to a voltage drop across the structure. The effect was first described in Ref.~ \cite{08} within the hydrodynamic approach, which is applicable  for  the systems   operating in the electron-electron  collision dominated regime.
The properly designed two-dimensional (2D)   plasmonic structures
yield superior detection of the THz radiation \cite{1n,3n,4n,5n,6n,8n,10n,11n,12n}.
THz detectors based on GaAs
\cite{010,0015,3,2,6,12,013,014,0101,99,ad1,ad2}, Si \cite{7,nep1}  and  GaN \cite{2,nep2}  FETs have already achieved  performance comparable to  or even exceeding that of commercial detectors, demonstrating
tunability \cite{08,09,010,0015,3,2,6,12,013,014,0101,99,ad1,ad2,7,nep1,nep2}, a
relatively low value of the noise equivalent power \cite{nep1,nep2},  a potential to detect signals with
very high modulation frequencies  (up to hundreds of GHz) \cite{me-mod}), and operation in  heterodyne and homodyne regimes \cite {13n,14n,15n,16n,17n,apl2017} with a very high responsivity.  Such  detectors
     can operate both at zero bias  current, with a minimum shot noise, and in the regime of a relatively large  drain-to-source current. In the latter case, the detection efficiency can be significantly improved due to the current-driven increase of nonlinear properties of the channel \cite{0015,99}.
  Also, as any nonlinear elements,  plasmonic detectors can operate as  frequency mixers or frequency multipliers  \cite{08}.
Using multi gate detectors  based on  the  ratchet effect (see Refs.~\cite{rachet1,rachet2,rachet3,rachet4,7n} and references therein) further improves the
detector performance.
%and enables helicity-sensitive response

       Recently, we studied  the THz homodyne detection   in the strongly non-perturbative, with respect to radiation power, regime \cite{apl2017} (see also previous  publications on homodyne detection   \cite{13n,14n,15n,16n,17n} and on the non-perturbative response \cite{22n,23n,24n}). We found the upper bound for the rectified response, which exceeds the conventional  perturbative response by orders of magnitude. Most importantly, we also demonstrated  that, apart from the extremely high sensitivity, this regime of operation
        allows for the direct measurements of the phase difference between a weak incoming signal and the local oscillator signal. In other words, the homodyne response encodes the information about the phase difference  between the two signals (see also \cite{14n}). This property enables using plasmonic detectors as THz and far infrared  spectrometers and interferometers.

 Here, we show that helicity driven response could also enable
   the application of the {\it resonant} plasmonic   detectors as tunable  THz   spectrometers and interferometers.
     Helicity-driven {\it non-resonant} effects were observed earlier  in Ref.~\cite{Ganichev} and explained  in \cite{romanov}.  They were predicted to be absent  for zero loading impedances \cite{romanov}.   Below, we demonstrate that in the resonant regime, the  intrinsic FET channel  shows the helicity-driven response.

   We start from recalling that  plasma waves in a gated two-dimensional structure biased  above threshold,
have a linear dispersion law $\omega(k) = s k,$ \cite{01} where
 \begin{equation}
s=\sqrt{\frac{eU_g}{m}} \label{s}
\end{equation}
is the  wave velocity,
 which along with the electron concentration in the channel,
\begin{equation}
N=\frac{CU_g}{e}, \label{grad}
\end{equation}
 is controlled by the gate-to-channel  swing $U_g$ counted from the  FET threshold  voltage [in Eq.~\eqref{grad} we assume that $U_g$ is positive and much larger than the thermal voltage].
  Here $C=\epsilon/4\pi d$ is the gate to channel capacitance per unit area, $e$ is the electron charge, $m$ is the
electron effective mass, $d$ is the gate-to-channel distance and
$\epsilon$ is the dielectric constant.
For low electron scattering rates, a  structure of a given length, $L$, acts
for plasma  waves as a resonant "cavity",  with   resonant  frequencies
\be \omega_N  =    \omega_0 (N+a),
\label{omegaN}
 \ee
where
 $ \omega_0=\pi s/L,$  $N=0,1,2,\ldots$, and
 $a$  is  numerical coefficient which depends on  BC: $a=1/2$ for  voltage fixed at the source  and  current fixed at the drain \cite{02} and $a=0$ for  voltage fixed both at the source and at the drain (for $a=0,$ the mode with $N=0$ corresponds to the Drude peak discussed at the end of the paper).
  In a short channel FET, the
oscillation frequency, $\omega_0/2\pi
$, can be tuned by $U_g$
 to be in THz range.

A quality factor of the cavity is given by
 $ \omega_0 \tau,$ where $\tau$ is the momentum relaxation
time.
Depending on the value of the   quality factor and excitation frequency,  transistor can operate in the resonant, $\omega \approx \omega_0 \gg 1/\tau, $ and non-resonant   regimes.  In the latter case,
the plasma oscillations are overdamped. The dramatic reduction of the device
sizes in the last decades has led to the development of the new
generations of FETs, which may have high quality factors. Such FETs
should demonstrate novel physics, specific for the ballistic
regime.  In particular, the response of such FET to an external radiation shows sharp resonances.
Typically, these resonances are well described by the {\it symmetric} Lorentz peaks  insensitive to the radiation polarization \cite{08}.  One of the purposes of this paper is to demonstrate that excitation by a circularly-polarized wave can result in the  helicity-sensitive  resonant response with an {\it asymmetric} line shape.
  We  assume that the
 electron-electron collisions are very fast turning the system into
 the hydrodynamics regime. The  hydrodynamic equations
describing a two-dimensional electronic fluid in FET channel  read
\begin{equation}
\label{eq1} \frac{\partial v}{\partial t} + v\frac{\partial
v}{\partial x} + \gamma {v} = - \frac{e}{m} \frac{\partial
U}{\partial x},
\end{equation}
\begin{equation}
\label{eq2} \frac{\partial U}{\partial t} + \frac{\partial \left(
{Uv} \right)}{\partial x} = 0,
\end{equation}
where     $v$ is  the velocity of the electronic fluid, and
$U$ is the local value of the gate-to-channel voltage, which, in the gradual channel approximation, is
is related to the local value of the electron
concentration  as
$
N(x)={CU(x)}/{e}
  $ \cite{02}.
  %This equation  is a simple generalization of Eq.~\eqref{grad}.
  The rate of the velocity relaxation is determined  by the inverse momentum relaxation time $\gamma=1/\tau.$
 Actually, there is also some  momentum-dependent contribution to the  relaxation rate, $\eta k^2 \sim \eta/L^2,  $ caused by the viscosity $\eta$ of the electron fluid \cite{02}.  Here, we neglect this contribution assuming that  $L^2 \gg \eta \tau.$
 We  also assume that channel is sufficiently wide and do not discuss the effects  related  to the friction of the viscous electron fluid  at the boundaries of the sample.

Eqs.~\eqref{eq1} and ~\eqref{eq2} require two BC,
which depend on the properties of contacts. In the first publications on the plasma-wave  non-linear detection
Ref.~\cite{02},  \cite{08} it was  assumed that the ac  voltage $U_a$  is  applied  at
the source side of the channel and the current flowing through the FET
   is fixed at the  drain  side of the channel.  Physically, this implies the infinite inductive loading impedance  on the drain side of the channel.
It was shown  \cite{08} that rectification of the
ac oscillations induces a  constant
source-to-drain voltage:
$V \propto U_a^2$  at low intensities of excitation \cite{08} and  $V \propto U_a$ at higher intensities \cite{apl2017,22n,23n,24n}.

Here, we consider the BC corresponding to a different physical situation \cite{Ganichev,romanov}.   We assume that the circularly polarized  radiation excites  a sample via two antennas coupled to the source and drain.  Hence,  ac signals  at the source and  drain have equal frequencies but can be  shifted by phase with the shift magnitude $\theta$  determined by antenna design and have different amplitudes.  Taking into account that the  radiation induces a dc voltage  drop $V$ across the sample,  we write the BC as follows
\BEA
&&U(0)= U_g+U_a \cos(\omega t + \theta),   \nonumber \\
&& U(L)= U_g+V+U_b \cos \omega t .
\label{bound1}
\EEA
In the case, when the device is excited by a circularly polarized wave,  $\theta  $
changes sign with changing the helicity of the polarization \cite{Ganichev,romanov}.  We focus on the helicity-driven effects, i.e. on the contribution to the current, which changes sign with replacing $\theta$ with $-\theta$  or, equivalently, $\omega$ with $-\omega$ (for definiteness we  put below $\omega>0$).     We  will derive general equation for the dc response  valid both in resonant and  non-resonant cases,  and find that in all cases helicity-dependent part of the response  is given by
 \be
J_{\rm hel} \propto \sin \theta
~  U_aU_b.
\label{hel}
\ee
 We  will   show that the coefficient in this equation  dramatically increases   in vicinity of plasmonic resonances \cite{comment0} as compared to  non-resonant case dicussed  in  Refs.~\cite{Ganichev, romanov}.

We first introduce the dimensionless variable
$n={(U-U_g)}/{U_g}$
 and search for the solution of
Eqs.~\eqref{eq1},\eqref{eq2} in the
following form
  \begin{align}
  &n = n_0(x) + \frac{1}{2}n_1(x)e^{-i\omega t} +
  \frac{1}{2}n_{-1}(x)e^{i\omega t} + ...,
  \label{5}
  \\
  &v = v_0(x) + \frac{1}{2}v_1(x)e^{-i\omega t} +
  \frac{1}{2}v_{-1}(x)e^{i\omega t} + \cdots,
  \label{6}
 \end{align}
where  $n_0(x) = \langle n(x,t) \rangle_t$ and $v_0=\langle v (x,t) \rangle_t$ are the time-averaged potential and velocity, respectively, and    $n_{ 1}$ and $v_{ 1}$    are small ($\propto U_{a,b}$)  radiation-induced  plasmonic oscillations.   In the absence of radiation, $n_0=0,v_0=0$, while in the presence of radiation they are quadratic  with respect to the wave amplitude  ($\propto U_{a,b}^2$).

 Substituting Eqs.~\eqref{5} and \eqref{6} into
Eqs.~\eqref{eq1}, \eqref{eq2} and averaging
over time we get
  \begin{equation}
  \frac{\partial}{\partial x}\left( \frac{v_0^2}{2} + \frac{v_1v_{-1}}{4} +
  s
  ^2 n_0\right) + \gamma v_0 = 0,
  \label{15}
  \end{equation}
  \begin{equation}
  \frac{\partial}{\partial x}\left[ (1+n_0)v_0 + \frac{n_1v_{-1} +
  n_{-1}v_1}{4}\right] = 0.
  \label{16}
  \end{equation}
 Since the voltage is fixed at the source, the BC at the source is $n_0(0)=0.$
 By solving these equations, one can find the radiation-induced voltage drop across the sample,
  $V=U_g [ n_0(L) -n_0(0)]=U_g  n_0(L) .$
  In  Eqs.~\eqref{15}, \eqref{16},  one can neglect
    small terms $\p v_0^2/\p x$ and  $\p (v_0 n_0)/\p x$ \cite{comment1}.
       For zero dc current, we find from Eq.~\ref{16}  $ v_0 =- ({n_1v_{-1} +
  n_{-1}v_1})/{4}.$ Substituting this equation into Eq.~\ref{15} we find
            \BEA
    &&\frac{V}{U_g}=\frac{1}{4 s^2} \left[   \gamma \int_0^L  dx (n_1 v_{-1}+ n_{-1} v_1) \right.
   \nonumber
    \\
    &&\left.+  v_1(0)v_{-1}(0)  - v_1(L)v_{-1}(L) \right].
    \label{dn}
    \EEA
  Next,  one  should  find $n_1$ and $v_1$ and substitute into Eq.~\eqref{dn}.
 One can see that the variations of $n_0$ and $v_0$ are small and can be neglected in quadratic in $U_{a,b}$ approximation provided  that  there is no dc current in the channel.
  Therefore, in   equations for $n_{ 1}$  and $v_{ 1}$ we can assume
 $n_0 = const$, $v_0 = const.$   Since   the dc current in the channel is zero, we put  $v_0=0.$   One can also  assume in these equations $n_0=0,$ since spatially independent concentration is  fully  controlled by $U_g.$
 Then, we get
 \BEA
 && (\gamma -i \omega ) v_1 +s^2\frac{\p n_1}{\p x}=0
 \label{dn1}
 \\
 \label{dv1}
 && -i\omega n_1 +\frac{\p v_1}{\p x} =0.
 \EEA
In the infinite system, the solutions of Eqs. ~\eqref{dn1}, \eqref{dv1} are harmonic plasma waves
  $n_1 , v_1 \sim e^{\pm ikx},$
  with
 \be
 k=\frac{\sqrt{\omega (\omega +i\gamma) }}{s} =\frac{\Omega+i\Gamma}{s}.
 \label{k}
  \ee
  Here,
  $\Omega=s(k+k^*)/2, ~ \Gamma=s(k-k^*)/2i$
  are the  plasma wave frequency and  damping, respectively.

    The solution of Eqs.~\eqref{dn1} and \eqref{dv1} in the  finite system of length $L$   with the  BC \eqref{bound1} reads
 \BEA
 &&
 n_1=\left(A e^{i k x}+B e^{-ikx}\right),
 \label{n1}
 \\
 &&
 v_1= \frac{\omega}{k} \left[A e^{i k x}-Be^{-ikx} \right],
 \label{v1}
 \EEA
 where
 \BEA
 A=\frac{ U_b e^{-i\theta_b}  - U_a e^{-i\theta_a}e^{- ikL}}{2i U_g  \sin(kL)} ,
  \\
  B=\frac{U_a e^{-i\theta_a} e^{ ikL} - U_b e^{-i\theta_b}}{2i U_g  \sin (kL)} .
  \EEA
    Substituting Eqs.~\eqref{v1} and \eqref{n1} into  Eq.~\eqref{dn}, we find
  \be
  V=  \frac{\omega}{\sqrt{\omega^2 +\gamma^2}}\frac{\alpha (U_a^2-U_b^2) +\beta U_a U_b \sin \theta}{4 U_g \left| \sin (kL)\right|^2},
  \label{response}
  \ee
  where
  \begin{align}
  & \alpha= \left(\! 1\!+\!\frac{\gamma \Omega\!}{ \Gamma \omega}\right) \sinh^2\!\left(\!\frac{\Gamma L}{s}\!\right)\!-\! \left( \! 1\!-\!\frac{\Gamma \gamma}{ \Omega \omega}\!\right) \sin^2\!\left(\!\frac{\Omega L}{s}\!\right),
  \\
  & \beta=8\sinh\!\left(\!\frac{\Gamma L}{s}\!\right)\sin\!\left(\!\frac{\Omega L}{s}\!\right).
    \end{align}

 Equation \eqref{response}  is valid for an arbitrary relation between $\omega,\omega_0$ and $\gamma.$ Different regimes of operation are illustrated in Fig.~\ref{fig1}. Below we discuss these regimes in detail.

\begin{figure}[ptb]
\centering
\includegraphics[width=0.40\textwidth]{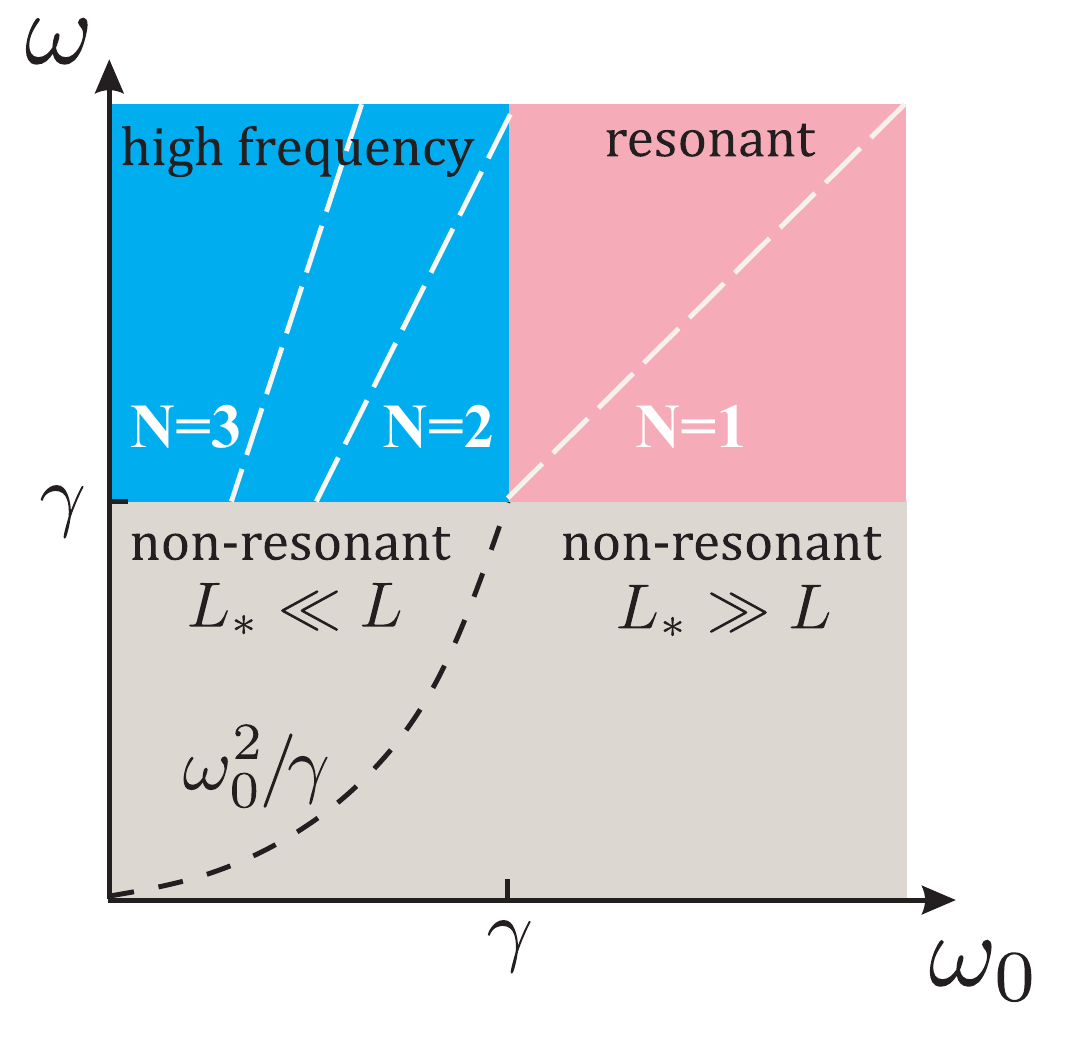}
\caption{Different regimes of detector operation.      }
\label{fig1}
\end{figure}

\subsubsection{Non-resonant case, $\omega \ll \gamma$ (grey area in Fig.~\ref{fig1}).}
 In this case,
 from Eq.~\eqref{k},  we find:
   $\Omega\approx \Gamma \approx  \sqrt{\omega \gamma/2},$
  which means that the plasma waves  are overdamped.
 It is convenient to  introduce the characteristic length \cite{08,99}
 \be
 L_*=  \frac{s\sqrt 2}{\sqrt{\omega \gamma}}=\frac{s}{\Gamma}.
 \ee
 The plasma excitations  exponentially decay at the scale $L_*$ from the edges of the sample to the bulk.
One can, therefore, consider two limiting cases of the long and  short samples:
\paragraph{{Long sample, $L\gg L_*$}} ($\omega \gg \omega_0^2/\gamma$).
 In this case, Eq.~\eqref{response} simplifies
     \be
    V=\frac{U_a^2-U_b^2 +16 U_aU_b e^{- L/L_*} \sin(L/L_*)(\omega/\gamma) \sin \theta }{4 U_g}.
   \label{hel-long}
    \ee
      The  helicity-sensitive term  is exponentially small. This is   because helicity-dependent contribution arises due to  the coupling between the source and drain, which is suppressed  in the long sample.
       \paragraph
      {{Short sample, $L\ll L_*$}} ($\omega \ll \omega_0^2/\gamma$ ).
                       In this  case,
              the response reads
       \be
    V=\frac{U_a^2-U_b^2 +4 U_aU_b (\omega/\gamma) \sin \theta }{4 U_g}.
   \label{hel-short}
    \ee
Let us compare Eqs.~\eqref{hel-long} and \eqref{hel-short} with the analytical results obtained in Ref.~\cite{romanov}, which was focused on the study of the non-resonant case.  Their analysis  demonstrated that a non-zero contribution to the heleicity-sensitive part of the response      appears only for non-zero loading impedance $Z.$ Since in our case $Z\equiv0,$ it seems that there is contradiction between the results. This contradiction is resolved by noticing that  in   Eqs.~\eqref{hel-long} and \eqref{hel-short} the helicity-driven contribution contains a  factor $\omega/\gamma$ which is small in the  non-resonant approximation. The non-resonant equations used in Ref.~\cite{romanov}    neglect such terms.
In other words,  the terms, which are  proportional to $ \sin \theta,$ appear in       Eqs.~\eqref{hel-long} and \eqref{hel-short} as corrections to pure  non-resonant approximation. It worth stressing  that  in a short sample  the  helicity-driven contribution  is  not exponentially small    and, therefore, can be   observed  experimentally.
%%%%%%%%%%%%%%%%%%%%%%%%%%
%%%%%%%%%%%%%%%%%%%%%%%%%

\begin{figure}[ptb]
\centering
\includegraphics[width=0.42\textwidth]{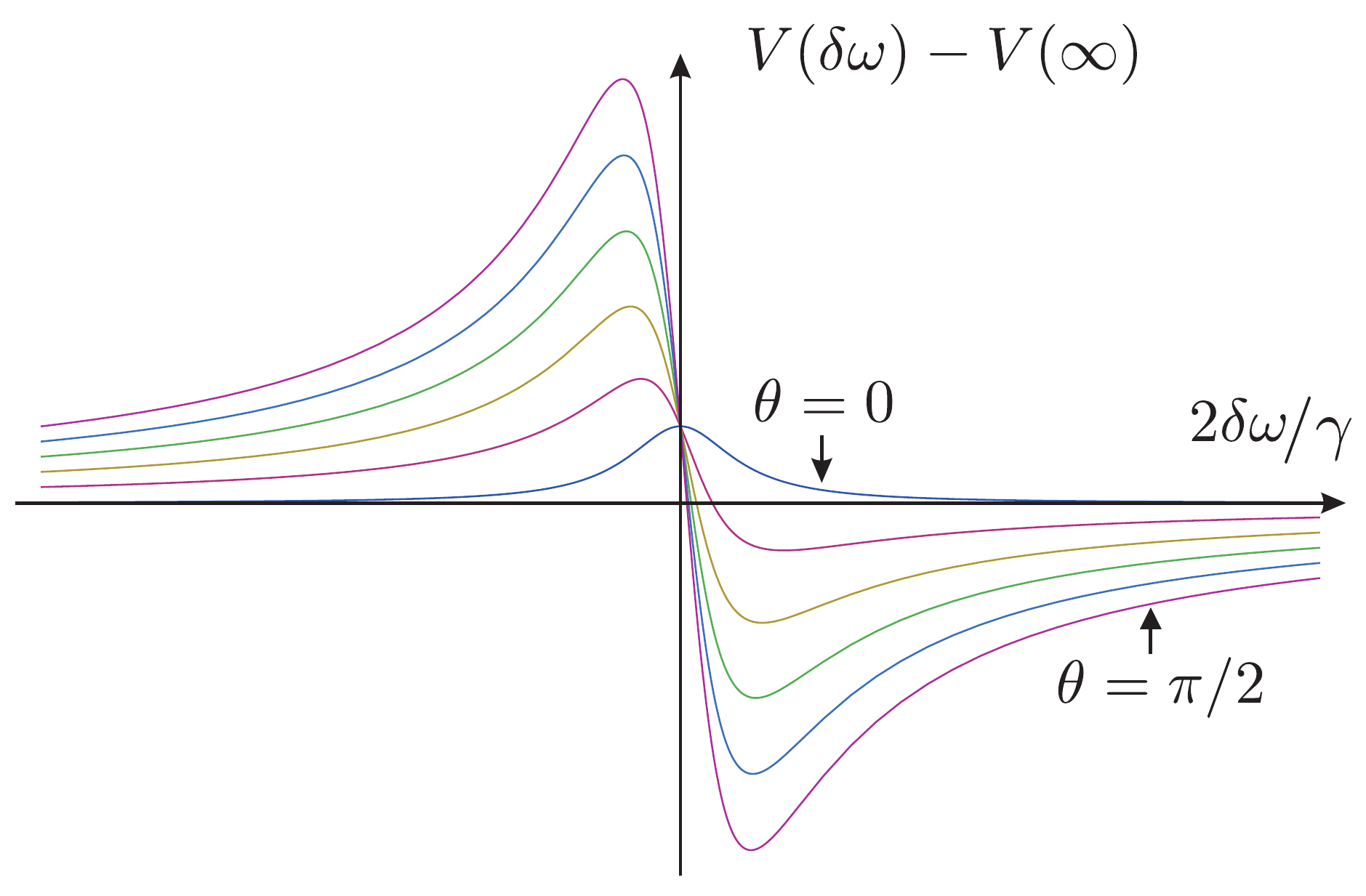}
\caption{Resonant dependence of response on the radiation frequency [Eq.~\eqref{V-dOmega}] at different
$\theta$ varying from $0$ to $\pi/2$ for  $U_aU_b/(U_a^2-U_b^2)=5$ and  $N=1.$ With increasing $\theta$ asymmetry of the resonance is enhanced due to increasing of the helicity-sensitive contribution.     }
\label{fig2}
\end{figure}
%%%%%%%%%%%%%%%%%%%%%%%%%%
\subsubsection{High frequency  case, $\omega_0 \ll \gamma \ll \omega $ (blue area in Fig.~\ref{fig1}).}
In this case,
$\Omega \approx \omega,  \Gamma \approx \gamma/2.$
 The response is given by
\be
V=\frac{3(U_a^2-U_b^2)}{4 U_g} +\frac{4U_aU_b e^{-\gamma L/2s} \sin(\pi \omega/\omega_0) \sin \theta }{U_g}.
\label{high omega}
\ee
Similar to Eq.~\eqref{hel-long}, the helicity-dependent contribution in Eq.~\eqref{high omega} is exponentially small, since $\gamma L/s \sim \gamma/\omega_0 \gg 1$. However, the frequency dependencies  of these equations are essentially different: linear in Eq.~\eqref{hel-long} and periodic in Eq.~\eqref{high omega}.

\subsubsection{Resonant case, $\omega \gg\gamma,~\omega_0 \gg\gamma $ (pink area in Fig.~\ref{fig1}) }
    Similar to the previous case, $\Omega \approx \omega,  \Gamma \approx \gamma/2.$    The response shows series of the sharp peaks at $\omega =\omega_N.$ In the vicinity of $N$-th resonance ($N\neq 0$), we find
 \be V(\delta \omega)\!=\!\frac{(U_a^2-U_b^2)(3\gamma^2/4-\delta\omega^2)
        \!+\!  {4 U_aU_b}(-1)^N \delta \omega \gamma \sin \theta}{{4 U_g}(\delta \omega^2+\gamma^2/4) }
       \label{V-dOmega}
       \ee
       where
       $\delta\omega =\omega-\omega_N$ and  resonant frequencies are given by  $\omega_N= \pi N s/L$ for BC Eq.~\eqref{bound1} [$a=0$ in Eq.~\eqref{omegaN}].

The most intriguing property of Eq.~\eqref{V-dOmega} is an asymmetrical  resonance  dependence on the frequency of the incoming radiation.  More specifically, the response is given by the sum of two  parts sharply peaked at $\delta\omega=0$: conventional, polarization-independent part, which obeys the symmetry  $\delta \omega \to -\delta \omega,$ and helicity-driven  part  which  changes sign under this operation. The latter increases with increasing the phase shift $\theta$. This property is illustrated in Fig.~\ref{fig2}, where response is shown for fixed $\gamma$ and  $U_{a,b}$ but for different phase shifts $0<\theta<\pi/2.$ As seen, asymmetrical part of the response increases with increasing $\theta.$  Hence, conventional and helicity-driven contributions can be easily separated  by measuring the frequency dependence of the response.

\subsubsection{Drude peak}
Finally, we consider in more detail what happens when $\omega_0 \gg \gamma  $ and  $\omega$ increases from small values $\omega \ll \gamma$   to relatively large value  $\omega_0 \gg \omega \gg \gamma$ (moving from grey to pink area in Fig.~\ref{fig1}). In other words, we consider response for $\omega \approx \omega_N $ with $N=0.$  Simple analysis of Eq.~\eqref{response} yields the peak of the width $\gamma$ (Drude peak)
\be V( \omega)\!=\!\frac{(U_a^2-U_b^2)(\gamma^2-\omega^2)
        \!+\!  {4 U_aU_b}  \omega \gamma \sin \theta}{{4 U_g}( \omega^2+\gamma^2) }.
       \label{V-dOmegaN=0}
       \ee
For $\omega \ll \gamma$ this equation simplifies to Eq.~\eqref{hel-short}.

Comparing Eqs.~\eqref{V-dOmega} and \eqref{V-dOmegaN=0}, we find that the Drude peak  is very similar to the  plasmonic resonances. However, symmetrical and  asymmetrical parts of the Drude peak are, respectively, 3 and 4 times smaller.

 To conclude,
  the theory of  nonlinear resonant plasmonic  response of the  gated 2D  electron gas  subjected to  THz radiation with a given helicity shows that the  helicity-driven contribution dramatically increases in the vicinity of the plasmonic resonances.  This contribution  is a harmonic  function of the phase shift  and shows  an asymmetric dependence on the excitation frequency in the vicinity of  the resonances. Hence it can  be easily separated from conventional symmetric contribution, which is not sensitive to radiation polarization.  Helicity-sensitive contribution can also be observed in the non-resonant regime. Although it is small  in this case as compared to polarization insensitive part of the response, it has easily identifiable   frequency and phase dependence  and, therefore, can be experimentally separated.

The work of M. S. S.  was supported by the U.S. Army Research Laboratory through the Collaborative Research Alliance
%(CRA)
for Multi-Scale Modeling of Electronic Materials
%(MSME)
and by the Office of the Naval Research (Project Monitoe Dr. Paul Maki). The work of V. Yu. K. was supported by Russian Science Foundation (grant No. 16-42-01035). The work of I.V.G. was supported by the Foundation for the advancement of theoretical physics “BASIS”.


\begin{thebibliography}{99}
\bibitem{comment0} Since we consider here resonant case, we neglect smooth dependence of $U_a$ and $U_b$ on frequency which might arise due to frequency dependence of the loading impedances.
\bibitem{comment1} These  terms can be safely  neglected  keeping only quadratic order with repect to $U_{a,b},$ and considering zero   dc current in the channel.
        For discussion of a  more general situation see   Refs. ~\cite{99,13n,22n,23n}.
\end{thebibliography}
\end{document}